\documentclass[twocolumn,showpacs,preprintnumbers,amsmath,amssymb,superscriptaddress,prb]{revtex4}
\usepackage{graphicx}% Include figure files
\usepackage{dcolumn}% Align table columns on decimal point
\usepackage{bm}% bold math
\usepackage{color} % \color{red} \color{black}
\usepackage{ulem}

\begin{document}

\title{Kondo resonance in PrTi$_2$Al$_{20}$: Photoemission spectroscopy and single-impurity Anderson model calculations}

\author{M.~Matsunami}
 \altaffiliation[Present address: ]{UVSOR Facility, Institute for Molecular Science, Okazaki 444-8585, Japan; matunami@ims.ac.jp}
\affiliation{Institute for Solid State Physics, University of Tokyo, Kashiwa, Chiba 277-8581, Japan}
\affiliation{RIKEN SPring-8 Center, Sayo-cho, Sayo-gun, Hyogo 679-5148, Japan}
\author{M.~Taguchi}
\author{A.~Chainani}
\author{R.~Eguchi}
 \altaffiliation[Present address: ]{Graduate School of Natural Science and Technology, Okayama University, Okayama 700-8530, Japan.}
\author{M.~Oura}
\affiliation{RIKEN SPring-8 Center, Sayo-cho, Sayo-gun, Hyogo 679-5148, Japan}
\author{A.~Sakai}
\author{S.~Nakatsuji}
\affiliation{Institute for Solid State Physics, University of Tokyo, Kashiwa, Chiba 277-8581, Japan}
\author{S.~Shin}
\affiliation{Institute for Solid State Physics, University of Tokyo, Kashiwa, Chiba 277-8581, Japan}
\affiliation{RIKEN SPring-8 Center, Sayo-cho, Sayo-gun, Hyogo 679-5148, Japan}

\date{\today}% It is always \today, today,

\begin{abstract} 
The Kondo resonance at the Fermi level is well established for the electronic structure of Ce (4$f^1$ electron) and Yb (4$f^1$ hole)-based systems. 
In this work, we report complementary experimental and theoretical studies on the Kondo resonance in the Pr-based 4$f^2$ system, PrTi$_2$Al$_{20}$. 
Using Pr\,3$d$-4$f$ resonant photoemission spectroscopy and single impurity Anderson model (SIAM) calculations including the full multiplets of Pr ions, we show that a 4$f^2$ system can also give rise to a Kondo resonance at the Fermi level. 
The Kondo resonance peak is experimentally observed through a final-state-multiplet dependent resonance and is reproduced with properly tuned hybridization strength in SIAM calculations. 
\end{abstract}


\maketitle

The Kondo effect represents a typical manifestation of many-body phenomena and its experimental signature is a logarithmic temperature ($T$) dependence in the electrical resistivity, $\rho(T)$$\sim-\log$\,$T$. \cite{Hewson} 
It arises from the spin-flip scattering of conduction ($c$) electrons by a local magnetic impurity, and leads to an energetically narrow feature in the density of states (DOS) at the Fermi level ($E_{\rm F}$), often referred to as Abrikosov-Suhl or Kondo resonance. 
This feature can be observed even in the case of a periodic array of magnetic ions such as Ce-based intermetallics exhibiting heavy-fermion behavior at low $T$. 
The electronic structure and the Kondo resonance peak of Ce- and Yb-based systems (characterized by a formal 4$f^1$ configuration, Ce$^{3+}$: 4$f^1$ electron, Yb$^{3+}$: 4$f^1$ hole), have been successfully described within the framework of the single-impurity Anderson model (SIAM) based on $c$ states hybridizing with localized $f$ states. \cite{GS} 
Subsequent studies have also shown the importance of lattice effects based on the periodic Anderson model. \cite{YbIr2Si2} 
On the other hand, although the importance of hybridization effects has been nicely clarified in the electronic structure of $f^n$ systems, \cite{PrTM_Kucherenko1,EuNi2P2} the Kondo resonance at $E_{\rm F}$ has been elusive even for an $f^2$ configuration (Pr$^{3+}$: 4$f^2$ electron). \cite{PrFe4P12_Yamasaki} 
While spectroscopic experiments for U-based 5$f^2$ electron systems have shown the Kondo resonance in analogy to Ce-based systems, \cite{YUPd3_Liu} a combined validation of Kondo resonance in experiments and SIAM calculations including the full multiplets of the $f^2$ configuration has not been established. 
In this Brief Report, we carry out experimental and theoretical verification of the $f^2$ Kondo resonance at $E_{\rm F}$ in PrTi$_2$Al$_{20}$, a recently discovered 4$f^2$ Kondo system. \cite{PrTi2Al20_Basic}

Among Pr-based compounds, the $-\log$\,$T$ dependence in $\rho(T)$ has been observed so far in systems PrSn$_3$, \cite{PrSn3_1} PrFe$_4$P$_{12}$, \cite{PrFe4P12_Sato} Pr$_2$Ir$_2$O$_7$, \cite{Pr2Ir2O7_Nakatsuji} and very recently in Pr$M$$_2$Al$_{20}$ ($M$ = Ti and V). \cite{PrTi2Al20_Basic} 
Pr$M$$_2$Al$_{20}$ crystallizes in the cubic CeCr$_2$Al$_{20}$ structure, characterized by a cagelike coordination of Al ions surrounding Pr ions. 
The $\rho(T)$ in PrTi$_2$Al$_{20}$ decreases monotonically with cooling, and shows two shoulders at $\sim$50 and $\sim$2\,K. 
The magnetic contribution in $\rho(T)$, which is obtained by subtracting $\rho(T)$ of LaTi$_2$Al$_{20}$, shows a maximum at $\sim$50\,K and $-\log$\,$T$ dependence at higher $T$. 
The anomaly at $\sim$2\,K is also observed in the specific heat but not so clearly in the magnetic susceptibility, suggesting a nonmagnetic phase transition such as a quadrupolar ordering.

In this study, we address the Kondo resonance in the 4$f^2$ configuration using photoemission spectroscopy (PES) of PrTi$_2$Al$_{20}$. 
In order to directly probe the Kondo resonance feature with PES, we need to specifically enhance the Pr\,4$f$ component with enough bulk sensitivity for the following reasons. 
In the case of low concentration of the rare-earth element such as PrTi$_2$Al$_{20}$, it is difficult with use of normal (nonresonant) PES to extract the 4$f$ electron signal from the total DOS in the presence of many other orbital contributions. 
In addition, it is well known that 4$f$ electrons in the surface have a greater tendency to localize, \cite{Bulk} acting to prevent the Kondo behavior. 
Accordingly, we have selected Pr\,3$d$-4$f$ resonant PES as a bulk sensitive probe. 
The obtained on-resonance (Pr\,4$f$) spectra for PrTi$_2$Al$_{20}$ reflect the strong $c$-$f$ hybridization and reveal the existence of the Kondo resonance peak just at $E_{\rm F}$. 
No such feature is observed in elemental Pr metal, used as a prototypical system of localized 4$f$ electrons. 
Furthermore, we analyze the 4$f$ spectra within the framework of SIAM including the full multiplets of Pr ions and discuss the relationship between a Kondo resonance and a valence instability in 4$f^2$ configuration.

Single crystals of PrTi$_2$Al$_{20}$ were grown by the Al self-flux method. \cite{PrTi2Al20_Basic} 
Clean sample surfaces were obtained by fracturing $in~situ$. 
Pr metal was prepared as thin films by $in~situ$ evaporation. 
Pr\,3$d$-4$f$ x-ray absorption spectroscopy (XAS) and resonant PES were performed at the undulator beamline BL17SU in SPring-8. 
XAS spectra were recorded using the total electron yield method. 
PES spectra were measured using a hemispherical electron analyzer, Scienta SES-2002. \cite{BL17PES} 
The total energy resolution of PES was set to 250 and 100\,meV for the measurements of the overall valence-band region and near-$E_{\rm F}$ region, respectively. 
$E_{\rm F}$ of the samples was referenced to that of evaporated Au film. 
PES spectra were normalized to the incident photon flux, measured as the drain current at the focusing mirror. 
The $T$ was 25\,K and the vacuum was below 4$\times$10$^{-8}$\,Pa during all the measurements.

%%%%%%%%%%%%%%%%%%%%  FIG.1  %%%%%%%%%%%%%%%%%%%%
\begin{figure}[t]
\begin{center}
\includegraphics[width=0.45\textwidth]{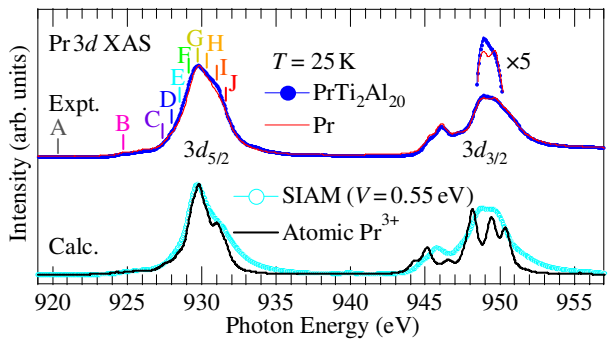}
\caption{
(Color online) 
Pr\,3$d$-4$f$ XAS spectra of PrTi$_2$Al$_{20}$ and Pr metal in comparison with those calculated by the SIAM with the hybridization strength $V$=0.55\,eV and the atomic multiplet model for the Pr$^{3+}$ ion. 
The labels $A-J$ indicate the selected photon energies at which the resonant PES spectra in Figs. 2 and 3 are measured. 
} 
\end{center}
\end{figure}
%%%%%%%%%%%%%%%%%%%%%%%%%%%%%%%%%%%%%%%%%%%%%%%%%

Figure~1 shows the Pr\,3$d$-4$f$ XAS spectra of PrTi$_2$Al$_{20}$ and Pr metal in comparison with those calculated by the SIAM with the hybridization strength ($V$) of 0.55\,eV and the atomic multiplet model for Pr$^{3+}$ ion. 
Details of the calculations are described later. 
The SIAM result gives good correspondence to both spectra. 
The minimal difference between both materials can be identified in terms of the peak broadness in the 3$d_{5/2}$ region and the peak shape at 948-951\,eV in the 3$d_{3/2}$ region. 
The variation from Pr metal to PrTi$_2$Al$_{20}$ follows a trend from atomic multiplet model to SIAM, suggesting a stronger hybridization in PrTi$_2$Al$_{20}$.  
Further details of the electronic structure are discussed below on the basis of Pr\,3$d$-4$f$ resonant PES, which probes more clearly the difference in the Pr\,4$f$ states between both materials.

%%%%%%%%%%%%%%%%%%%%  FIG.2  %%%%%%%%%%%%%%%%%%%%
\begin{figure}[t]
\begin{center}
\includegraphics[width=0.45\textwidth]{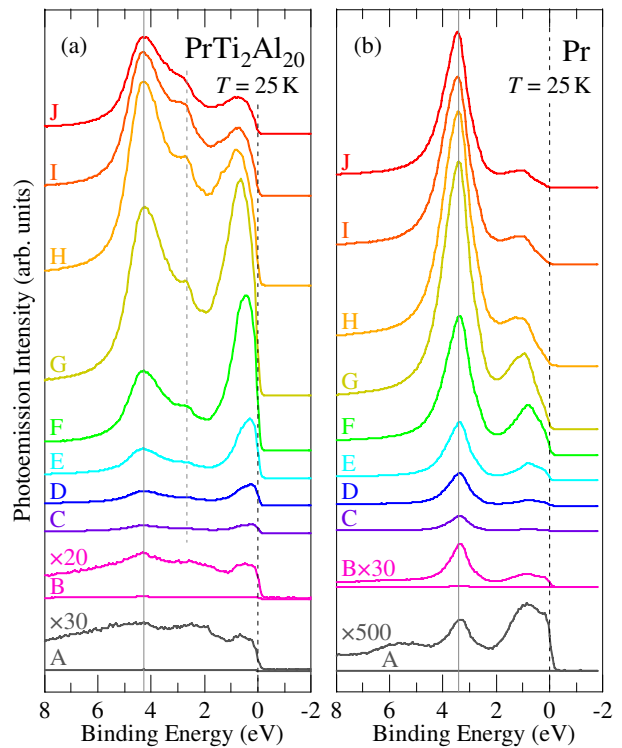}
\caption{
(Color online) 
Pr\,3$d$-4$f$ resonant PES spectra of PrTi$_2$Al$_{20}$ (a) and Pr metal (b) measured at the incident energies labeled on XAS spectra in Fig.~1. 
The solid and dashed gray lines indicate the main peak position of the $f^1$ final state and the lower-$E_{\rm B}$ shoulder, respectively. 
}
\end{center}
\end{figure}
%%%%%%%%%%%%%%%%%%%%%%%%%%%%%%%%%%%%%%%%%%%%%%%%%

Figure~2 shows the Pr\,3$d$-4$f$ resonant PES spectra of PrTi$_2$Al$_{20}$ (a) and Pr metal (b). 
For Pr metal, both on-resonance ($G$) and off-resonance ($A$) spectra are qualitatively similar to the previous data. \cite{Pr_Hufner} 
Taking into account the photoionization cross sections \cite{crosssec} and the elemental composition, the off-resonance spectrum is expected to be dominated by Ti\,3$d$, Al\,3$s$, and 3$p$ states in PrTi$_2$Al$_{20}$ and Pr\,5$d$ and 4$f$ states in Pr metal. 
Due to the resonance enhancement of the Pr\,4$f$ state, the evolution of two prominent features at 3-5\,eV of the binding energy ($E_{\rm B}$) and near $E_{\rm F}$ ($E_{\rm B}$=0-2\,eV) is confirmed for both materials, which can be attributed to the $f^1$ and $f^2$ final states, respectively. \cite{Pr_Hufner} 
In contrast to the unshifted $f^1$ final state for both materials as indicated by the solid lines in Fig.~2, the peak maximum of $f^2$ final state shifts in $E_{\rm B}$, depending on the incident photon energy. 
This is caused by the multiplet effect as discussed later, and is not due to the Auger effect. 
The weight of the $f^2$ final state relative to that of the $f^1$ final state, which gives a measure of the degree of $f$-electron localization, is much stronger in PrTi$_2$Al$_{20}$, suggesting the existence of itinerant $f$ electrons due to the stronger hybridization effect. 
Also, the positions and the separation of two features are quite different for the two materials. 
As another remarkable difference, an additional peak or shoulder can be seen at $\sim$2.7\,eV for on-resonance spectra only in PrTi$_2$Al$_{20}$. 
A similar feature has been reported in the filled skutterudite compounds. \cite{PrFe4P12_Yamasaki}

%%%%%%%%%%%%%%%%%%%%  FIG.3  %%%%%%%%%%%%%%%%%%%%
\begin{figure}[t]
\begin{center}
\includegraphics[width=0.45\textwidth]{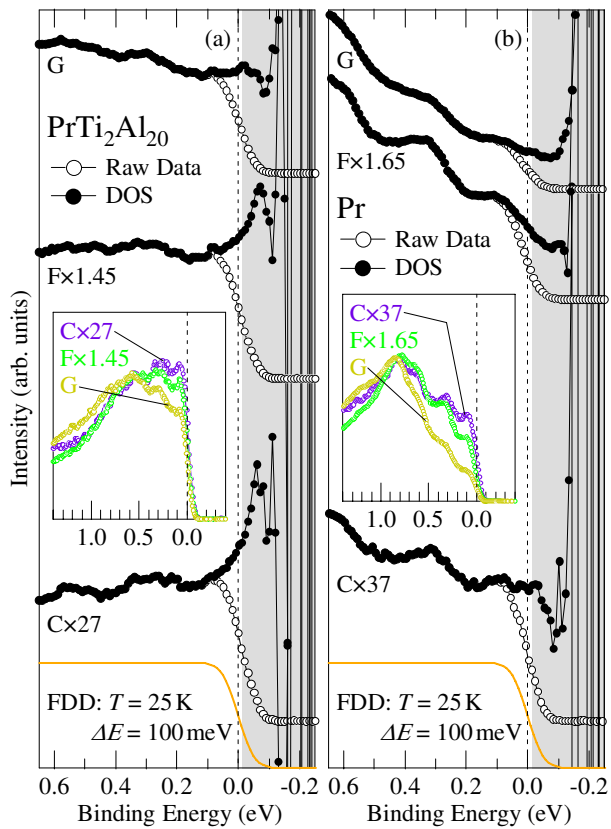}
\caption{
(Color online) 
High-resolution Pr\,3$d$-4$f$ resonant PES spectra measured at the incident energies $C$, $F$ and $G$ labeled on XAS spectra in Fig.~1 for PrTi$_2$Al$_{20}$ (a) and Pr metal (b), alongwith the corresponding spectral DOS obtained by dividing the resolution-convoluted FDD function. 
The shaded area shows the region of $E_{\rm B}<-5k_{\rm B}T$. 
} 
\end{center}
\end{figure}
%%%%%%%%%%%%%%%%%%%%%%%%%%%%%%%%%%%%%%%%%%%%%%%%%

Next, we focus on the $f^2$ final state. 
Figures~3(a) and (b) show the high-resolution spectra measured at the incident energies $C$, $F$, and $G$ for PrTi$_2$Al$_{20}$ and Pr metal, respectively, alongwith the corresponding spectral DOS obtained by dividing the resolution-convoluted Fermi-Dirac distribution (FDD) function. 
The observed fine structures are attributed to the multiplets of the $f^2$ final state, \cite{Pr_Hufner} as also confirmed by our SIAM calculations. 
Crucially for the spectra $C$ and $F$ in PrTi$_2$Al$_{20}$, the midpoint of the leading edge in the PES spectra locates slightly above $E_{\rm F}$ as is typically seen in many Ce systems, \cite{Ce_Joyce,Ce_Garnier,CeCu2Si2_Reinert,CeOs4Sb12} and in contrast to the case of Pr metal for which it locates at $E_{\rm F}$. 
This observation leads to the characteristic feature in the spectral DOS, in which the sharp rise at $E_{\rm F}$ toward the unoccupied side is obtained for PrTi$_2$Al$_{20}$ in contrast to the relatively flat DOS around $E_{\rm F}$ for Pr metal. 
In analogy to the case of Ce systems, this sharp rise should be identified as the ``tail'' of the Kondo resonance peak, responsible for the Kondo effect in $\rho(T)$. 
Note that the spectrum $G$ for PrTi$_2$Al$_{20}$, which is measured at the peak top in XAS, hardly shows an enhancement of the tail near $E_{\rm F}$, even though the total weight of the $f^2$ final state is most strongly enhanced. 
It suggests a multiplet dependent resonance in the $f^2$ final state, which moves to higher $E_{\rm B}$ with increasing incident energy as seen in the inset of Fig.~3(a). 
Such behavior, which is similar in Pr metal as shown in the inset of Fig.~3(b), can be described by considering the intermediate-state multiplet effect in the resonant process (equivalent to the XAS final state). 
The enhancement of individual final-state multiplet depends on the corresponding intermediate-state multiplet. 
For PrTi$_2$Al$_{20}$, the Kondo resonance peak derived from the lowest-$E_{\rm B}$ final-state multiplet resonates at the incident energy of absorption prepeak. 
Then, it gradually gets buried in the higher-$E_{\rm B}$ multiplets with increasing incident energy toward the absorption peak maximum ($G$). 
Indeed, the slope of Kondo resonance peak at $E_{\rm F}$ is sharper in spectrum $C$ than in $F$. 
The main part of resonance in spectrum $F$ moves to slightly higher $E_{\rm B}$. 
Thus, the on-resonance spectra obtained at the absorption prepeak energies such as $C$-$E$ correspond to the genuine Pr\,4$f$ spectra reflecting the Kondo resonance at $E_{\rm F}$.

%%%%%%%%%%%%%%%%%%%%  FIG.4  %%%%%%%%%%%%%%%%%%%%
\begin{figure}[t]
\begin{center}
\includegraphics[width=0.45\textwidth,clip]{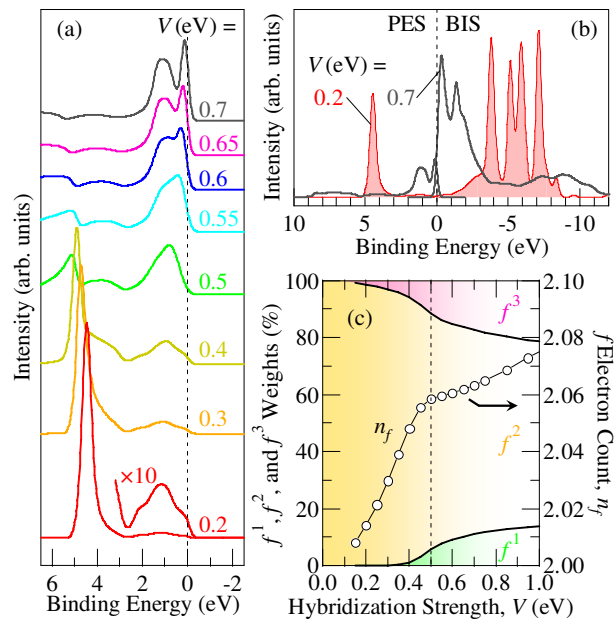}
\caption{
(Color online) 
The SIAM simulations for Pr ($f^2$) systems. 
(a) Calculated Pr\,4$f$ PES spectra as a function of $V$. 
(b) PES and BIS spectra. 
(c) The $V$ dependence of the distribution of 4$f$ weights (left axis) and the 4$f$ electron count $n_f$ (right axis). 
} 
\end{center}
\end{figure}
%%%%%%%%%%%%%%%%%%%%%%%%%%%%%%%%%%%%%%%%%%%%%%%%%

To reproduce the Pr\,4$f$ spectra of PrTi$_2$Al$_{20}$ and Pr metal, we have performed SIAM calculations as a function of $V$. 
The basis set consisting of $4f^1$, $4f^2$, and $4f^3$ configurations is used to describe the ground state. 
The $V$ between the $4f$ and $c$ band depends on the $c$ band energy. 
For the $c$ band states, we use a Lorentzian profile. \cite{Patthey} 
The Slater integrals and spin-orbit coupling constants are calculated by the Hartree-Fock method with relativistic corrections and are reduced to 85\% and 96\%, respectively. \cite{Cowan} 
To take into account the configuration dependent hybridization, $V$ is reduced by a factor $R_{\rm C}$ (=0.8) in the presence of a core hole and enhanced by a factor $1/R_{\rm V}$ ($R_{\rm V}=0.95$) in the presence of an extra 4$f$ electron. \cite{Rc&Rv} 
Details of the model calculations are standard and are described in earlier work. \cite{GS,Groot} 
In order to obtain the best fit of calculations to the experimental data of PrTi$_2$Al$_{20}$, we needed to use the following parameter values: 
bare 4$f$ energy $\varepsilon_f$=$-$3.2\,eV, 
on-site Coulomb repulsion $U_{ff}$=7\,eV (standard in Pr-based compounds \cite{PrTM_Kucherenko1}), 
core-hole potential $U_{fc}$=10\,eV, and 
$c$ bandwidth $W$=1.6\,eV. 
In addition to the PES spectra, the bremsstrahlung isochromat (BIS) spectra shown in Fig.~4(b) and the XAS spectra shown in Fig.~1 have also been calculated using the same parameters.  
It should be emphasized that the full multiplet structures of Pr ions are included in our calculations. 
It was found to play an essential role in analyzing the Pr\,4$f$ spectra, since they show pronounced multiplet features.

Figure~4(a) shows the calculated Pr\,4$f$ PES spectra as a function of $V$. 
With increasing $V$, the peak at 3-4\,eV corresponding to the $f^1$ final state shifts toward higher $E_{\rm B}$ together with splitting off to the lower-$E_{\rm B}$ shoulder. 
Then, the $f^2$ final state at 0-2\,eV gradually develops to compensate the suppressed weight of the $f^1$ final state. 
Moreover, the intense part in the $f^2$ final state moves closer to $E_{\rm F}$, leading to the Kondo resonance peak. 
Such behavior is consistent with the spectral variation from Pr metal to PrTi$_2$Al$_{20}$. 
The calculation corresponding to $\sim$0.55\,eV shows the best match to the experimental data of PrTi$_2$Al$_{20}$ in terms of the multiplet lineshape of the $f^2$ final state and the splitting between the $f^1$ final state and its lower-$E_{\rm B}$ shoulder. 
This demonstrates that a fractional valence ($n_f$$\sim$2.06 at $V$\,=\,0.55\,eV) is indeed realized in PrTi$_2$Al$_{20}$, which should be closely related to the Kondo effect in $\rho(T)$. \cite{PrTi2Al20_Basic} 
A valence instability of the Pr ion was also reported in the dilute alloy Pr (1.4$\%$) in Pd metal, which exhibits a Kondo effect in $\rho(T)$ and a fractional valence ($\sim$+3.1). \cite{PrinPd}

The evolution of the Kondo resonance peak is more clearly seen in BIS spectra as shown in Fig.~4(b). 
Here, the PES and BIS spectral weights are normalized to be $n_f$ and $N_f$$-$$n_f$, respectively, where $n_f$ is the calculated $f$ electron count and $N_f$ (=14 in this case) is the degeneracy of the $f$ electron. 
For BIS, the multiple peaks around $-2$ to $-9$\,eV for $V$=0.2\,eV and the near-$E_{\rm F}$ peaks around $0\sim-3$\,eV for $V$=0.7\,eV are attributed to the $f^3$ ($f^2 \to f^3$) and the $f^2$ ($f^1 \to f^2$) final states, respectively. 
With increasing $V$, the spectral weight transfers from $f^3$ to $f^2$ final state, while in the case of PES, it transfers from the $f^1$ ($f^2 \to f^1$) to the $f^2$ ($f^3 \to f^2$) final state, giving rise to an effective incoherent-to-coherent crossover. 
A major part of the $f^2$ final state including the Kondo resonance peak exists in the BIS side, providing the peak top of Kondo resonance in the unoccupied DOS, as in the case of Ce systems. \cite{GS} 
This is also consistent with our experimental findings.

Of particular interest is that $n_f$ indicates a plateau above $V$=0.5\,eV as shown by the dashed line in Fig.~4(c). 
This indicates that the initial growth in $f^3$ weight gets compensated by the $f^1$ weight on increasing $V$ above 0.5\,eV. 
Since this compensation is due to the fluctuation between an electron-in ($f^3$) and an electron-out ($f^1$) of the $f$ shell with respect to the $f^2$ ground state through the $c$-$f$ hybridization, it is related with the evolution of the Kondo resonance. 
This fact suggests that the $n_f$ can be stabilized by the Kondo resonance. 
The $V$$\sim$0.5\,eV provides a boundary for distinguishing the Pr\,4$f$ electronic states near $E_{\rm F}$ between the Kondo resonance regime and the nearly localized regime, which PrTi$_2$Al$_{20}$ and Pr metal belong to, respectively. 
It is important that the $n_f$ ($\sim$2.06) in PrTi$_2$Al$_{20}$ shows only a little deviation from Pr$^{3+}$ and a little difference between PrTi$_2$Al$_{20}$ and Pr metal in spite of the large spectral variation in PES. 
A similar $n_f$ stabilization with varying $V$ has been reported in the study of Pu metal with 5$f^5$ ground state. \cite{Pu_PES} 
Further experimental and theoretical investigations on other 4$f$ or 5$f$ electron systems are necessary to establish the relationship between the $n_f$ stabilization and the Kondo resonance for $f^n$ systems in general.

To summarize, we have performed Pr\,3$d$-4$f$ resonant PES on PrTi$_2$Al$_{20}$ and Pr metal. 
The intense $f^2$ final state is observed for Pr\,4$f$ spectra in PrTi$_2$Al$_{20}$, reflecting the stronger $V$ and the Kondo resonance behavior. 
The $f^2$ final state shows a multiplet-dependent resonance. 
The Kondo resonance peak at $E_{\rm F}$ in PrTi$_2$Al$_{20}$ is identified in the spectra measured at absorption prepeak energies, while it is suppressed in Pr metal. 
The spectral variation between PrTi$_2$Al$_{20}$ and Pr metal can be reproduced by the $V$-dependent SIAM calculations including full multiplets of Pr ions. 
The SIAM calculations suggest that $n_f$ is stabilized by the Kondo resonance.

This work is partially supported by Grant-in-Aid for Scientific Research (Grant No. 21684019) from JSPS, and by Grant-in-Aid for Scientific Research on Innovative Areas ``Heavy Electrons'' from MEXT, Japan.


\end{document}